\documentstyle[aps,prl,multicol,epsf]{revtex}
\begin{document}

%\draft command makes pacs numbers print

\draft

\title{Ideal glass-glass transitions and logarithmic decay of correlations in a simple system.}

%repeat the \author\address pair as needed

\author{L. Fabbian, W. G\"otze
\footnote{{
Permanent address: Physik-Department, Technische Universit\"at 
M\"unchen, 85747 Garching, Germany}}
, F. Sciortino, P. Tartaglia, F. Thiery}
\address{
Dipartimento di Fisica, Universit\`a di Roma La Sapienza \\and
Istituto Nazionale di Fisica della Materia, 
Unit\`a di Roma {\it La Sapienza} \\
Piazzale Aldo Moro 2,
I-00185 Roma,Italy}

\date{\today}

\maketitle

\begin{abstract}
We calculate the ideal-glass-transition line for adhesive hard spheres
in the temperature-volume-fraction plane within the framework of the 
mode-coupling theory. We find two intersecting lines, controlled by the
hard-core and the adhesive part of the potential respectively, giving
rise to two different mechanisms for structural arrest.  In the glass
region we identify the presence of a glass-glass-transition line
ending in a cusp bifurcation which causes, even in the close by liquid
region, a logarithmic decay of correlations.
\end{abstract}

\pacs{PACS numbers: 64.70.Pf, 82.70.Dd}

\begin{multicols}{2}
The crossover from a liquid to an amorphous solid, observed near the
calorimetric glass transition temperature $T_{g}$, exhibits as a
precursor phenomenon an anomalous dynamics, called glassy dynamics.
Its evolution is connected with a critical temperature $T_c$ above
$T_{g}$.  It has been studied extensively in the recent literature of
the glass-transition problem, both
experimentally\cite{cummins,tolle,petry,vanmegen,underwood},
numerically\cite{kob,water} and
theoretically\cite{bengtzelius,report}.  Experiments around $T_c$ have
been interpreted in the frame of the mode-coupling theory (MCT) for
structural relaxation. MCT deals primarily with closed equations of motion for
the normalized density-fluctuation-correlation functions $\Phi_q(t)$
for wavevector moduli $q$. The equilibrium structure enters as input
in these equations via the static structure factor $S_q$.  The theory
explains $T_c$ as a glass-transition singularity resulting as a
bifurcation phenomenon for the self-trapping problem of density
fluctuations.  Below $T_{c}$ the interaction of density fluctuations
leads to arrest in a disordered array, characterized by a Debye-Waller
factor $f_q>0$.  Near the transition, the MCT equations can be solved
by asymptotic expansions. Beyond the initial transient dynamics,
correlation functions are predicted to decay with a time fractal
toward a plateau value $f_q^{c}$, the critical Debye-Waller factor.
Above
$T_c$ the correlations decay from $f_q^c$ to zero, and this is
the MCT interpretation of the $\alpha$-process of the classical 
literature of glassy dynamics\cite{angell}. The initial part of this 
decay is another time fractal, called   von Schweidler's law. 
The values of the fractal exponents are controlled by the so-called 
exponent parameter $\lambda \le 1$, which depends solely on $S_q$.
For details and citations of the original literature the reader is
referred to the review in Ref.\cite{report}.  A number of tests of MCT
results against data, among them the ones in
Refs.\cite{cummins,tolle,petry,vanmegen,underwood,kob,water,nauroth,semi},
demonstrates that this theory treats reasonably the evolution of
structural relaxation in some systems. 

The MCT bifurcations are caused by a nondegenerate eigenvalue
of a certain stability matrix to approach unity from
below\cite{leshouches}. Therefore the bifurcation scenario for
$f_q$ is that known for the zeroes of a polynomial
as induced by changes of the polynomial's
coefficients\cite{arnold}. The generic case for a change due to a
single control parameter is a fold
bifurcation\cite{arnold}, i.e. the
coalescence of two real zeroes to a degenerate one. 
In all the microscopic models for liquid-glass
transitions studied so far, only the fold bifurcation singularity had been identified. The
next more complicated scenario is the cusp bifurcation, 
equivalent to
the coalescence of three real zeroes.  Generically, this case can be
obtained only if two control parameters are
varied\cite{arnold}.  The MCT-bifurcation dynamics for this case
has been worked out in a leading order  asymptotic expansion\cite{gosj}. It is
drastically different from the one for a fold bifurcation in the sense
that relaxation stretching is much more pronounced. In particular
there appear logarithmic decay laws, i.e. $1/f$-noise spectra. 
Cusp bifurcations in MCT are endpoints of transition lines characterized by
$\lambda=1$, and typically they are located near intersections
of bifurcation lines\cite{leshouches}. 
Sj\"ogren\cite{sjogren} identified
some sets of polymer dielectric-loss spectra
which could not be fitted
by the fold-scenario but were compatible with a cusp bifurcation.
However, in the data used in Ref.~\cite{sjogren} the temperature was 
varied as the only
control parameter, and therefore a complete test of the
cusp-bifurcation scenario was not possible.

In this paper we want to demonstrate that a cusp bifurcation near the
intersection of two fold-bifurcation lines is possible in a simple but
realistic model, if the two conventional control parameters
temperature $T$ and density $n$ are varied. With this aim we study an extension
of the hard sphere system (HSS), the sticky hard spheres system (SHSS) 
introduced
by Baxter\cite{baxter68} and used extensively, in
particular in colloid physics\cite{review,verduin}.  
The HSS is the
archetype of a simple system and its only control parameter is the
packing fraction $\phi=\pi n \sigma^3/6$, with $\sigma$ denoting the
hard sphere diameter.  It has been studied by dynamic light scattering
in its realization as a certain colloidal suspension. The system
exhibits an ideal glass transition at some critical packing fraction
$\phi_{c}$\cite{vanmegen}.  This is a sharp transition of the fully
equilibrated sample from an ergodic liquid to a nonergodic amorphous
solid. The glassy dynamics of the colloid exhibits the same scaling
laws and $f_q$ anomaly as found for the more conventional 
systems\cite{cummins,tolle,petry,kob,water,nauroth,semi}.
The measured correlators can be explained quantitatively by the 
first-principle calculation produced by MCT\cite{underwood}.  The SHSS
potential is the limiting form of a square well,
infinitely narrow and infinitely deep\cite{menon}. It is defined by the
interparticle potential
$\beta V(r) =\infty$ for $r \le \sigma$, 
$ln\left(12 \tau \Delta / (\sigma+\Delta)\right)$ 
for $\sigma<r\le \sigma+\Delta$, 
and 0 for $r > \sigma+\Delta $, 
in the limit $\Delta \rightarrow 0$.
Here $\beta = \left({k_B T}\right)^{-1}$, $k_B$ is 
Boltzmann's constant, 
and $\tau$  is the adhesiveness parameter
which plays the role of a dimensionless temperature. For  $\tau \rightarrow \infty$ 
one recovers the HSS.
The Ornstein-Zernike equation 
$S_q=1/(1-n c_q)$
with the Percus-Yevick closure has been 
solved analytically by Baxter and we use his expressions for 
$S_q$, as a function of $\tau$ and $\phi$\cite{baxter68} as input of 
this work.
The crucial point is that the interaction 
exhibits two features, which favour arrest of density fluctuations for 
two quite different reasons. One is a strong short-ranged repulsion.  It is 
the mechanism leading to the cage effect for liquid dynamics and 
thereby to arrest driven by density fluctuations near the structure 
factor peak position.  The other one is a short-ranged attraction.  It 
leads via the usual mean-field mechanism to a softening of the density 
fluctuations, in particular of those for small wave vectors.  It is 
the mechanism which leads to a liquid-gas transition 
at low values of $\tau$.

The MCT equations of motions are\cite{report}
\begin{equation} 
\ddot{\Phi}_q(t)+\nu_q \dot{\Phi}_q(t)+\Omega_q^2 \Phi_q(t)+
\Omega_q^2 \int_0^t ds\,m_q(t-s) \dot{\Phi}_q(s)=0.
\label{eq:dyn}
\end{equation}
\noindent
Here $\Omega_q = \sqrt{q v /S_q}$, with $v$ denoting the thermal velocity,
is an effective phonon-dispersion law and $\nu_q=\nu_1 q^2$ denotes a damping constant (we use $\nu_1=1$).
The kernel $m_q$ is given as $m_q(t) = {\cal F}_q [\Phi_k(t)]$, where the mode-coupling
functional ${\cal F}_q$ is determined by the structure factor:
\begin{equation}
{\cal F}_q [f_k] = {1 \over 2} \int{ {{d^3k} \over 
(2 \pi )^3} V_{\vec{q},\vec{k}} f_k f_{|\vec{q}-\vec{k}|}},
\label{eq:mq}
\end{equation}
\begin{eqnarray}
V_{\vec{q},\vec{k}} \equiv  S_q S_k S_{|\vec{q}-\vec{k}|} {n \over {q^4}}
\left[ {\vec{q}} \cdot 
\vec{k}\,{c_k} +\vec{q} \cdot  
(\vec{q}-\vec{k})\,{{c_{|\vec{q}-\vec{k}|}} }
 \right]^2.
\label{eq:v}
\end{eqnarray}
\noindent
We numerically solved Eq.~(\ref{eq:dyn}) on a grid of 400
equally spaced $q$ values extending up to $q\sigma=72$\cite{tesi}.
$f_q$ is obtained by an iterative solution of the bifurcation equation
\begin{equation}
{f_q / (1-f_q)}={\cal F}_q[f_k].
\label{eq:fq}
\end{equation}

\begin{figure}[ht]
\epsfxsize=8cm\epsfysize=7.2cm\epsfbox{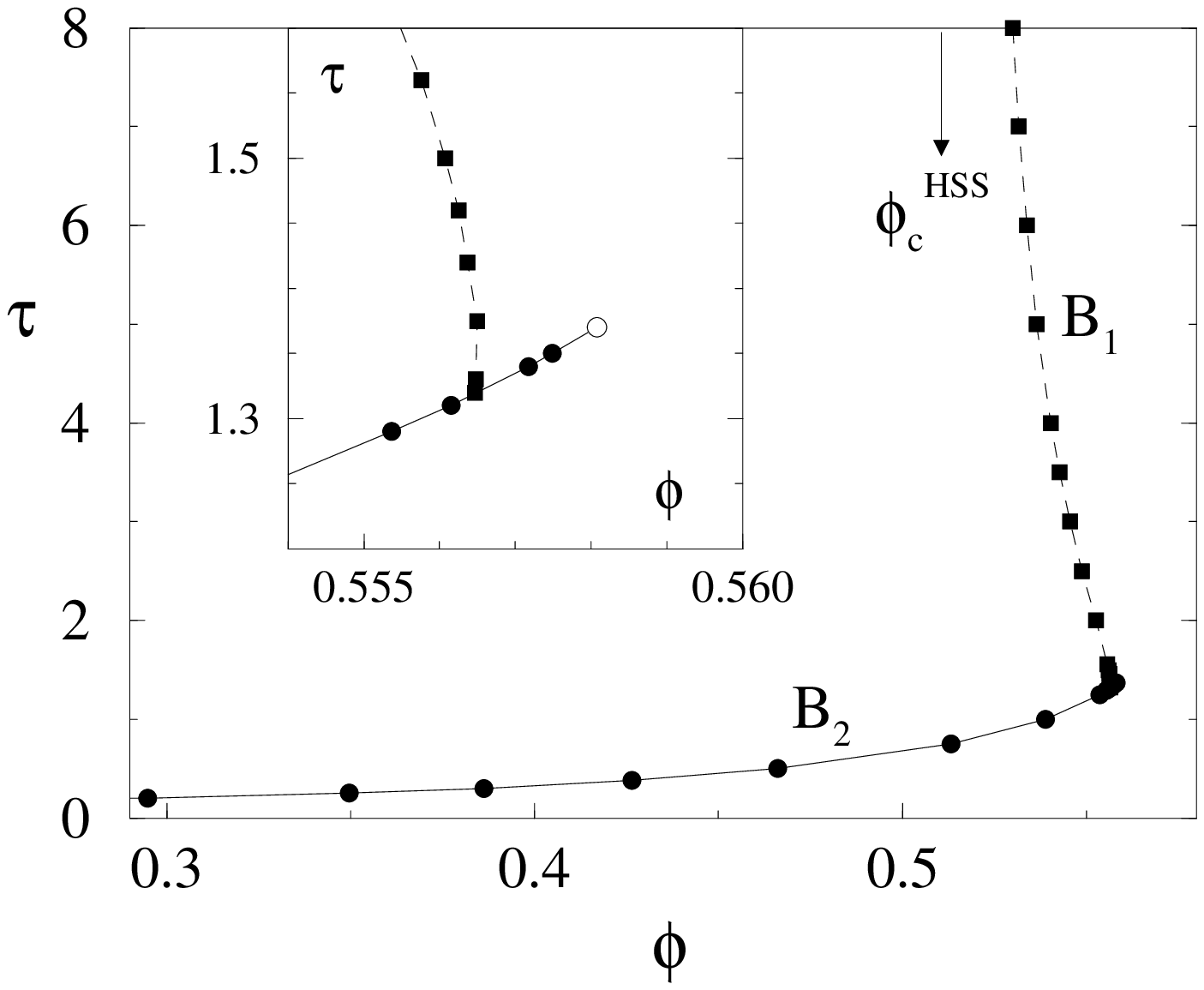}
\caption{
Ideal glass-transition lines for the Baxter model (squares and
dots). $B_1$ (dashed) is the high temperature branch of the
glass-transition line and $B_2$ (full) is the low temperature one. The
arrow indicates the critical packing fraction of the hard sphere
system.  The open circle denotes the cusp-point.
}
\label{fig:1}
\end{figure}

\begin{figure}[ht]
\epsfxsize=8cm\epsfysize=17.cm\epsfbox{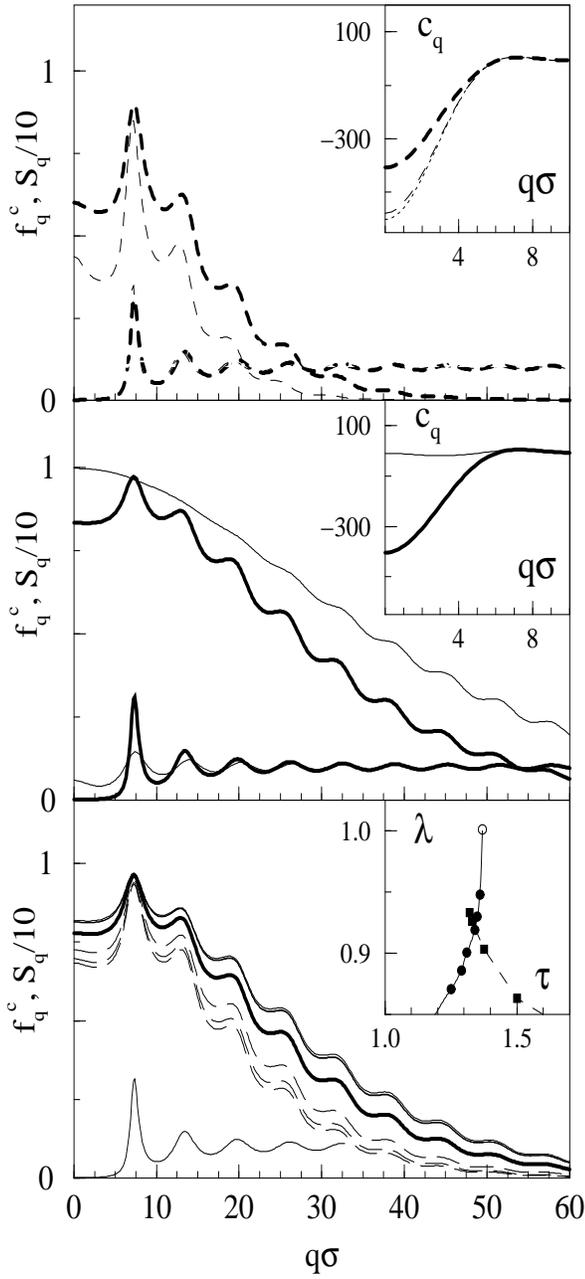}
\caption{
Structure factors $S_q/10$, direct correlation functions $c_q$ and
critical Debye-Waller factors $f_q^c$ along $B_1$ (top), $B_2$ (middle) and 
along the glass-glass transition line (bottom).
Top: 
$\tau_c = 8$ , $\phi_c = 0.530$ (light dashed lines) and 
$\tau_c = 1.32$, $\phi_c=0.522$ (heavy dashed lines). 
The dotted line in the inset is $c_q$ at $\phi_c^{HSS} = 0.516$ 
for $\tau=\infty$.
Middle:
$\tau_c=0.2$, $\phi_c=0.295$ (light full lines) and $\tau_c=1.31$,
$\phi_c=0.5561$ (heavy full lines).
Bottom:
The full (dashed) lines refer from top to bottom (bottom to top) to 
$\tau=1.34, 1.35, 1.36$ respectively, on the $B_2$ ($B_1$) side of the 
glass-glass transition line. The heavy full line is $f_q$ for the cusp point
$\tau=1.37$, $\phi=0.55807$. The lower $S_q$ refers to the same states.
The inset shows the exponent parameter $\lambda$ along $B_1$ (dashed) and 
$B_2$ (full).
}
\label{fig:2}
\end{figure}

For $\tau \rightarrow \infty$ we
recover the known HSS result, for which the ideal glass transition
at $\phi_{c}^{HSS} \sim 0.516$ \cite{bengtzelius} is lead by the increase of $-c_q$ and
$S_{q}$ with $\phi$.  For $\phi \ge \phi_c^{HSS}$, the particles are
trapped in cages formed by their neighbors. On decreasing $\tau$
attractive forces become relevant; nearest neighbor pairs are closer and
therefore holes in the cage are produced, destabilizing the
glass. A decrease of $\tau$ has to be compensated by an increase of
$\phi$ and thus the $\tau_c$ versus $\phi_c$ 
high-temperature-transition line has to bend away from the $\phi=\phi_c^{HSS}$
value. This effect is brought out by the MCT solutions as shown by the
dashed lines $B_1$ in Fig.~\ref{fig:1}.  The destabilization of the glass on
decreasing $\tau$ is related, within MCT, to the increase of pair
correlations which weaken the screened potential $-c_q$ and decrease
the value of $S_{q_{max}}$, as shown in Fig.~\ref{fig:2}-top.  Moving
along $B_1$ in Fig.~\ref{fig:1} by decreasing
$\tau$ implies an increase of the attraction and thus of the particle localization compared to the
HSS case.  Consequently, $f_q^c$
increases and spreads out further in the $q$-domain as demonstrated in
Fig.~\ref{fig:2}-top.

The increase of $c_q$ and $S_q$ at small $q$ becomes increasingly
relevant on decreasing $\tau$ producing 
a different mechanism for structural arrest. This increase of the compressibility is a precursor of the liquid-gas critical point. 
The enhanced amplitudes
for density fluctuations increase their anharmonic interactions, and indeed
the solution of the MCT equations predict their spontaneous arrest on a
low-temperature-transition line $B_2$ in the $\tau$ vs $\phi$ plane, shown
in Fig.~\ref{fig:1}.
Increase of $\phi$ suppresses small $q$
fluctuations. But it also increases $-c_q$ and the first peak of $S_{q}$, as shown
by a comparison of results shown in Fig.~\ref{fig:2}-middle. 
Both effects stabilize the glass, since the
mode-coupling constants, Eq.~(\ref{eq:v}) increase. This explains the upward bending of
the full line in Fig.~\ref{fig:1}.  Since arrest for small $q$ is
favored on the $B_2$ branch, $f_q^c$
is much larger there than for the HSS. The
decrease of the compressibility upon increasing $\phi$ make
spontaneous arrest less easy and therefore $f_q^c$
decreases upon compressing. 
Both features are demonstrated in Fig.\ref{fig:2}-middle.

The explained interplay of attraction, and repulsion places the meeting
point $C^*$ 
of the two branches $B_1$
and $B_2$ to a packing fraction $\phi^* = 0.5565$, higher  than $\phi_c^{HSS}$,
and to $\tau^* = 1.320$. Thus
there is a reentry phenomenon if $\tau$ is lowered, e.g. for a
packing fraction between $0.52$ and $0.55$. At some $\tau$ above
$1.32$ the glass melts since the increase of
attraction-induced-pair-correlations destabilizes the cages for
particle arrest. At some lower $\tau$ the liquid freezes again into a
glass, because of the compressibility increase.  At $C^*$,
the $f_q$ calculated on the two different branches, 
 $f_q^{(1)}$ and $f_q^{(2)}$,  are
different, while the corresponding $S_q$ are
indistinguishable. Thus two types of glasses, differing in
their $f_q$ and in their
dynamical properties, can be obtained close to $C^*$. 
In the region where both  $f_q^{(1)}$  and $f_q^{(2)}$ are
solutions of the Eq.~(\ref{eq:fq}), the one with the larger $f_q$ is
the relevant solution.
This exemplifies the maximum theorem of MCT: if
there is a solution of Eq.~(\ref{eq:fq}), say $\tilde{f_q}$, the long
time limit $f_q$ at the same equilibrium state obeys
$f_q \ge \tilde{f_q}$\cite{leshouches}.  For our case it means that the
continuation of $f_q^{c(2)}$ takes over the role of $f_q$, i.e. 
$B_1$ stops at $C^*$. 
Therefore the two branches of the
transition line do not join smoothly at $C^*$. 
The $B_2$ branch
continues, as shown by the inset of Fig.~\ref{fig:1} until it reaches 
a cusp bifurcation as
 an endpoint of the transition line.
Indeed, we evaluate the
exponent parameter $\lambda$ along the $B_1$ and $B_2$ lines and
find, as shown in the inset of Fig.~\ref{fig:2}-bottom, that along the $B_2$ line  
$\lambda$ keeps on
increasing till it reaches unity for $\tau=1.37$ and $\phi=0.55807$.
The line between $C^*$ and the cusp
point is a line of glass-to-glass transitions. For the given equilibrium
structure for ($\tau, \phi$) on this line, there are two
possibilities for structural arrest. The one on high $\tau$
side is driven by the excluded volume  mechanism 
and the other one by the attraction-induced compressibility increase.
The latter mechanism leads to a larger
$f_q$ than the former. The endpoint is
characterized by their difference approaching zero, as
shown in Fig.~\ref{fig:2}-bottom.

\begin{figure}[ht]
\epsfxsize=8.5cm\epsfysize=7.2cm\epsfbox{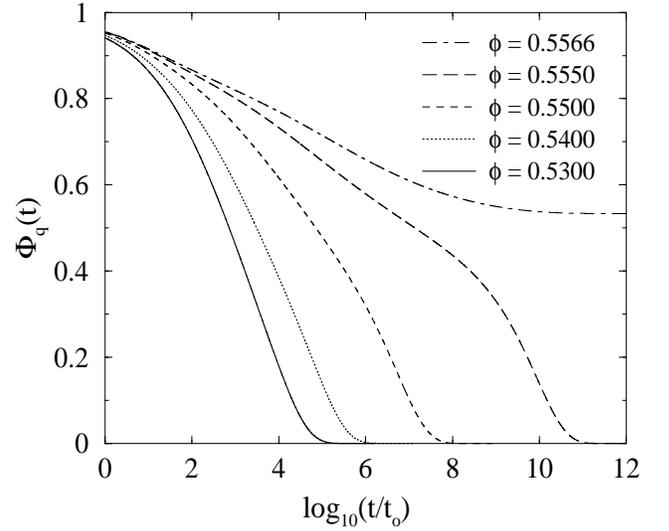}
\caption{
Density correlator for $q\sigma=14.4$ at $\tau=1.40$ and various $\phi$ as a function of $log_{10}(t/t_0)$ with $t_0=\sigma/v$.
}
\label{fig:5}
\end{figure}

In Fig.~\ref{fig:5} the glassy dynamics outside the transient is exhibited
for states in the reentry region. 
The dynamics is, up to a regular time scale, independent of the transient. In particular it is the same for a colloid, as for a conventional liquid for
which Eq.~(\ref{eq:dyn}) is formulated\cite{report}.
The uppermost curve 
corresponds to a point in the glass region and demonstrates
arrest near the $B_1$ branch, the others exhibit
stretched liquid relaxation to zero. Interestingly enough, the
exhibited anomalous dynamics is dominated by the cusp singularity, not
by the closer fold. Indeed even though a huge dynamical window is
considered, the known fold bifurcation pattern cannot be
recognized. There is no power-law decay towards the $f_q^{c(1)}$ nor
is there an $\alpha$-process obeying the superposition principle.  We
checked that the mentioned asymptotic fold-bifurcation features appear only 
after tuning $\phi$ much closer to the transition line, which implies the
extension of the dynamical window to even larger sizes. 
Instead, the cusp dynamics appears in the liquid region.
The
stretched relaxation in the three-decade window $0 \le log_{10}(t/t_0)
\le 3$ shows the approach towards the critical decay of the cusp\cite{gosj}. 
Then the correlators follow closely a logarithmic law in time. 
For $\phi=0.555$
it extends over 5 decades. Upon decreasing $\phi$ its range of
validity shrinks but even at $\phi=0.540$ it extends over two decades.
This logarithmic decay is the known cusp substitute for the start of a
fold $\alpha$-process. Let us emphasize that the above scenario does
not require a fine tuning of $\tau$. For $\tau=1.32$ the pattern is
similar to the one shown for $\tau=1.40$.

In summary, we have shown that the SHSS presents a peculiar structural
arrest dynamics at high packing fractions due to the competing
mechanism of hard core and attractive interactions. Two
differently sloped ideal-glass transition lines appear in the phase
diagram. In the region where these two lines meet the slow dynamics
changes from power-law to a logarithmic-law due to the influence of a
nearby cusp singularity.
In most tests of MCT the data had been compared with the universal
results obtained for the asymptotic dynamics near a fold
transition. But our results show that even a simple system can exhibit
glassy dynamics, which does not exhibit asymptotic laws within
accessible dynamical windows.  
It seems relevant 
%for a further assesment of MCT 
to test, by spectroscopy for colloidal systems
and by molecular-dynamics
simulations, whether specific systems exhibit the shown complex
dynamics and, in case, whether it is handled properly by the present MCT
calculation.

\narrowtext
\eject

\eject
\bigskip
\bigskip

\end{multicols}

\end{document}